\documentclass[aps,prb,amsmath,showpacs,twocolumn]{revtex4-1}

\usepackage{graphicx}
\usepackage{color}

\begin{document}

\title{Plasmon Dispersions in Simple Metals and Heusler Compounds}

\author{Steffen Kaltenborn}
\author{Hans Christian Schneider}
\email{hcsch@physik.uni-kl.de}
\affiliation{Physics Department and Research Center OPTIMAS, University of Kaiserslautern, 67663 Kaiserslautern, Germany}


\date{\today}

\begin{abstract}
We present a comprehensive study of plasmon dispersions in simple metals and Heusler compounds based on an accurate ab-initio evaluation of the momentum and frequency dependent dielectric function $\varepsilon(\vec{q},\omega)$ in the random-phase approximation. Using a momentum-dependent tetrahedron method for the computation of the dielectric function, we extract and analyze ``full'' and ``intraband'' plasmon dispersions: The ``full'' plasma dispersion is obtained by including all bands, the intraband plasma dispersion by including only intraband transitions. For the simple metals silver and aluminum, we show that the intraband plasmon dispersion has an unexpected \emph{downward} slope and is therefore markedly different from the results of an effective-mass electron-gas model and the full plasmon dispersion. For the two Heusler compounds $\mathrm{Co}_2\mathrm{Fe}\mathrm{Si}$ and $\mathrm{Co}_2\mathrm{Mn}\mathrm{Si}$, we present spectra for the dielectric function, their loss functions and plasmon dispersions. The latter exhibit the same negative intraband plasmon dispersion as found in the simple metals. We also discuss the influence of spin-mixing on the plasmon dispersion.
\end{abstract}

\maketitle

\section{Introduction}

Important characteristics of materials such as metals, half metals, and doped semiconductors are their opto-electronic properties due to the electrons in partially filled bands. These properties are determined by the dielectric function $\varepsilon(\vec{q},\omega)$, which is, in principle, a dynamical and wave-vector dependent quantity.~\cite{HaugKoch,Vignale,Mahan} In addition to the direct connection with optical ``constants'', the dielectric function also serves as input for calculations of various electronic properties, such as lifetimes or electronic dynamics.~\cite{Ladstadter,Zhukov} For materials with a ``simple'' band structure, e.g.,~metals and semiconductors, there exist approximate expressions for the dielectric function in the static and long-wavelength limit that are based on the electron-gas model for quasi-free electrons in the conduction band.~\cite{HaugKoch,Vignale,Mahan} This analytically tractable model is particularly useful for a qualitative understanding and can, in addition, be used as input in dynamical or lifetime calculations. On the other hand, starting as early as the 1950s, the dielectric function of ``simple'' materials has been measured~\cite{Ehrenreich,Shiles,Stahrenberg} and, in recent years, calculated \emph{ab initio} with impressive accuracy.~\cite{Draxl2,Onida,Ekardt,Lee,Schattke1,Schattke2} Both measurements and ab-initio calculations generally yield rather complex spectra that need to be interpreted with care, so that it would be helpful to have approximate results for the dielectric function based on the electron-gas model, even for materials where the band structure does not resemble a single band. For instance, one may want to study composition and crystal formation effects of Heusler compounds or alloys, by measuring their plasma frequency to have a single result that characterizes a particular growth condition.~\cite{Haeussler} These measured plasma frequencies may then be used to determine electronic densities and effective masses of a single-band electron-gas model. 

For the simple metals silver (Ag) and aluminum (Al) measurements~\cite{Ehrenreich,Piazza} and calculations~\cite{Draxl2,Eguiluz} showed that there is an influence on the plasmon dispersion from transitions \emph{between different bands}. This deviation from the single-band electron-gas model, even in the case of materials where it should apply best, leads to different views about what constitutes ``the'' plasma frequency: the intraband plasma frequency or the one corresponding to a peak in the loss function. Only very recently a negative plasmon dispersion for the layered compound $2H-\mathrm{NbSe_2}$ has been found and was ascribed to intraband transitions.~\cite{Silkin} In the present paper we complete the picture of different plasmon frequencies by taking a closer look at the dispersions of both candidates for the plasma frequency in simple metals and use the results to analyze the plasma behavior of the ``novel'' Heusler compounds $\mathrm{Co}_2\mathrm{Fe}\mathrm{Si}$ (CFS) and $\mathrm{Co}_2\mathrm{Mn}\mathrm{Si}$ (CMS).  

Our numerical approach uses a state-of-the-art calculation of the dielectric function in the random-phase approximation (RPA) from first principles. The electronic energies and matrix elements are obtained from a density-functional theory (DFT) calculation employing a full-potential linearized augmented plane wave (FP-LAPW) basis.~\cite{ELK} The evaluation of $\varepsilon(\vec{q},\omega)$ is done using an accurate wave-vector dependent linear tetrahedron method. We calculate the ``full'' dielectric function $\varepsilon(\vec{q},\omega)$ including transitions between all bands and the dielectric function including only intraband transitions from which we extract intraband and full plasma dispersions. We show that at finite wave vectors~$\vec{q}$, the two plasmon dispersions have very different slopes. In particular, the intraband plasmon dispersion curves downward, i.e., has a negative slope, which makes the difference between the two plasmons much more pronounced than it may seem from their $q\to0$ behavior in some materials, most notably aluminum. We stress that one needs to be aware of these differences between plasmon dispersions when using standard ab-initio codes, which usually compute the intraband plasma frequency. 

This paper is organized as follows. We give some details of our numerical approach in Sec.~\ref{numerics}. In Sec.~\ref{resultssimple} we present computed dielectric functions for aluminum and silver, and give a comprehensive discussion of the ``full'' and ``intraband'' plasma dispersions that are extracted from the dielectric functions. We then present a similar analysis of the dielectric functions and the plasma dispersions for the Heusler compounds in Sec.~\ref{resultsheusler}. Our results are summarized in Sec.~\ref{conclusions}.


\section{Numerical Method\label{numerics}}

We start by calculating the wave-vector dependent RPA dielectric function~$\varepsilon(\vec{q},\omega)$ based on electronic states and energies obtained from DFT calculations,~\cite{Draxl,Zhukov}
\begin{align}
\varepsilon(\vec{q},\omega) & =1-V_{q}\sum_{\mu\nu\vec{k}}
\big|B^{\mu\nu}_{\vec{k}\vec{q}}\big|^{2}
\frac{f_{\vec{k}}^{\nu}-f_{\vec{k}+\vec{q}}^{\mu}} {\hbar\omega+\epsilon_{\vec{k}}^{\nu}-\epsilon_{\vec{k}+\vec{q}}^{\mu}+i\hbar\gamma},
\label{eps}
\end{align}
where the limit $\hbar\gamma\to 0$ is understood. Here and in the following band indices are denoted by $\mu$, $\nu$, and wave vectors by $\vec{k}$, $\vec{q}$. The DFT energies and wave functions are $\epsilon_{\vec{k}}^{\mu}$ and $\psi_{\vec{k}}^{\mu}$, respectively, the $T=0$\,K occupation numbers $f_{\vec{k}}^{\mu}$, and the overlap matrix elements are defined by $B^{\mu\nu}_{\vec{k}\vec{q}}=\langle \psi_{\vec{k}+\vec{q}}^{\mu}|e^{i\vec{q}\cdot\vec{r}}|\psi_{\vec{k}}^{\nu}\rangle$, where we have neglected local field effects. Further, $V_{q}=e^2/(\varepsilon_{0}q^2)$ is the Fourier transformed Coulomb potential.


It is customary in state-of-the-art evaluations of the complex dielectric function to use a finite value of $\hbar\gamma$, which is usually either treated as an unavoidable parameter and/or chosen in accordance with experimental results.~\cite{Draxl,Zhukov,Tang} In the latter case, it is usually taken to be identical to the broadening of the Drude peak at $\omega=0$. We avoid the introduction of this parameter in the evaluation of~\eqref{eps} by first computing the imaginary part $\varepsilon_2\equiv\Im \varepsilon(\vec{q},\omega)$ in the limit $\hbar\gamma\to 0$,
\begin{equation}
\varepsilon_{2}=\pi V_{q} \sum_{\mu\nu\vec{k}}\big| B^{\mu\nu}_{\vec{k}\vec{q}} \big|^{2}\big(f_{\vec{k}}^{\nu}-f_{\vec{k}+\vec{q}}^{\mu}\big)
\delta\big(\hbar\omega+\epsilon_{\vec{k}}^{\nu}-\epsilon_{\vec{k}+\vec{q}}^{\mu}\big),
\label{eq:eps_2}
\end{equation}
with a linear tetrahedron method.~\cite{Eyert,MacDonald} This numerical method evaluates the integrand, which is only known on the numerical grid but contains a singular $\delta$ function, using a three-dimensional interpolation between the discrete $\vec{k}$-points in the first Brillouin zone (1.~BZ).~\cite{Eyert,MacDonald} On the finite grid, 8 $\vec{k}$-points form the edges of a parallelepiped, which is then split into 6 tetrahedra of equal size. The energy conservation $\hbar\omega+\epsilon_{\vec{k}}^{\nu}-\epsilon_{\vec{k}+\vec{q}}^{\mu}$ can be fulfilled by interpolating linearly between the edges of each tetrahedron. We have implemented this method, which is usually formulated for $q=0$ quantities such as plasma frequencies,~\cite{Draxl} $\varepsilon_2(\omega)$, reflectivities,~\cite{Schattke1,Schattke2} or the density of states,~\cite{Riedinger} for any finite $\vec{q}$ (and band combination $\mu$ and $\nu$) in order to obtain a numerically accurate result for $\varepsilon(\vec{q},\omega)$. The real part of the dielectric function, $\varepsilon_1\equiv \Re \varepsilon(\vec{q},\omega)$, is then found via the Kramers-Kronig relation (see, e.g., Ref.~\onlinecite{HaugKoch}).
The matrix elements $B_{\vec{k}\vec{q}}^{\mu\nu}$, energies $\epsilon_{\vec{k}}^\mu$ and distribution functions $f^{\mu}_{\vec{k}}$ have been computed in the framework of the density functional theory using the full-potential linearized augmented plane wave ELK code.~\cite{ELK} The calculation is in the same spirit as those presented in Refs.~\onlinecite{Puschnig,Ladstadter,Draxl}.

\begin{figure}
\includegraphics[width=0.40\textwidth]{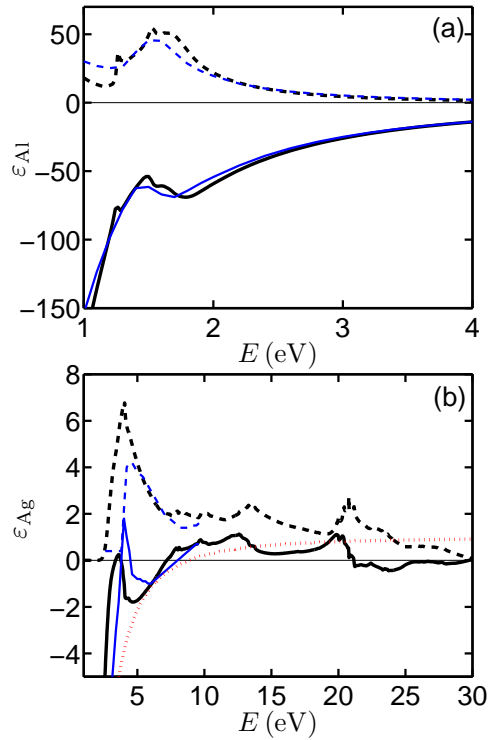}
\caption{\label{fig:eps}(Color online) Computed real (black solid line) and imaginary (black dashed line) part of the dielectric function $\varepsilon(\vec{q},\omega)$ for $\left|\vec{q}\right| = 0.44\,\mathrm{nm}^{-1}$. (a) Results for Al in comparison to optical measurements (thin blue lines).\cite{CRC,Shiles} (b) Results for Ag in comparison to optical measurements~\cite{Stahrenberg} (thin blue lines). The red dotted line shows the \emph{intraband} $\Re \varepsilon(\vec{q},\omega)$ whose zero determines the intraband plasmon resonance.}
\end{figure}

\section{Simple Metals\label{resultssimple}}

\subsection{Dielectric Function and Loss Function}

\begin{figure}
\includegraphics[width=0.40\textwidth]{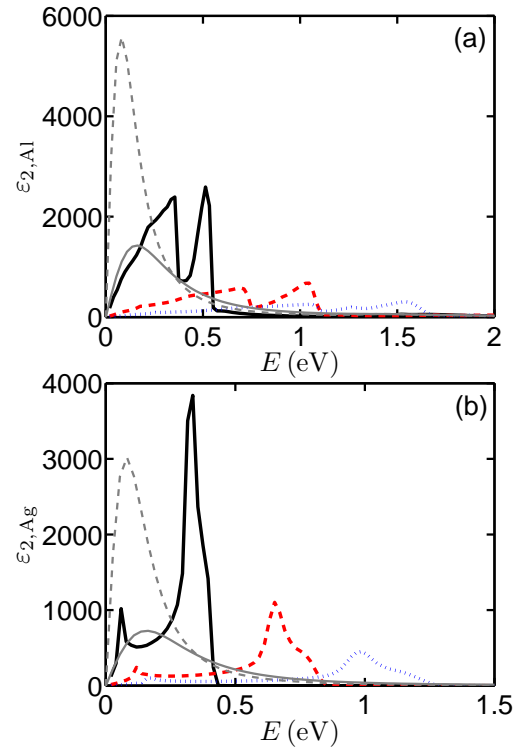}
\caption{\label{fig:eps2}(Color online) Imaginary part of the dielectric function of (a) aluminum and (b) silver for small energies and different $\vec{q}$. The black solid lines correspond to $|\vec{q}| = 0.44\,\mathrm{nm}^{-1}$ ($|\vec{q}| = 0.44\,\mathrm{nm}^{-1}$), the red dashed lines to $|\vec{q}| = 0.88\,\mathrm{nm}^{-1}$ ($|\vec{q}| = 0.87\,\mathrm{nm}^{-1}$) and the blue dotted lines to $|\vec{q}| = 1.32\,\mathrm{nm}^{-1}$ ($\left|\vec{q}\right| = 1.31\,\mathrm{nm}^{-1}$) in the aluminum (silver) calculation. Also shown are $q=0$ dielectric functions obtained directly from the ELK code for two different broadenings $\hbar\gamma = 0.01$\,Ha (thin grey line) and $\hbar\gamma=0.005$\,Ha (grey thin-dashed line).} 
\end{figure}

In Fig.~\ref{fig:eps} we present numerical results for the complex dielectric function $\varepsilon(\vec{q},\omega)$ of aluminum and silver because its dependence on $\vec{q}$ and $\omega$ forms the basis of the main results of this paper. We also provide a comparison with optical measurements \cite{CRC,Stahrenberg,Shiles} in order to check the accuracy of the calculation in the frequency range of interest. We included the first 20 (30) bands in the aluminum (silver) calculation and we used $61\times61\times61$ $\vec{k}$-points in the full Brillouin Zone. When comparing to the measured results one needs to keep in mind that these correspond to $q \approx 0$ (optical limit). We have to use a small but finite wave vector in the numerical evaluation of~\eqref{eps}. One therefore cannot avoid a discrepancy around $\omega=0$. Intraband transitions with vanishing momentum transfer $(q\to0)$ lead the Drude peak and consequently to a strong increase of $\varepsilon_{2}(0,\omega)$ for energies below~1\,eV, whereas our finite-$q$ results lead to a double-peak structure in this energy range, cf.~Fig.~\ref{fig:eps2}. Bearing this in mind we find a good agreement with Ref.~\onlinecite{Shiles} (cf.~also Ref.~\onlinecite{Lee}) in Fig.~\ref{fig:eps}(a) for Al in the range of $1\,\mathrm{eV}-4\,\mathrm{eV}$ and for $q = 0.44\,\mathrm{nm}^{-1}$. 

Figure \ref{fig:eps}(b) shows the computed dielectric function of silver for the same $\vec{q}$ on an wider energy range. The imaginary part again increases strongly for small energies due to intraband scattering processes (cf.~Fig.~\ref{fig:eps2}(b)), while the influence of interband transitions can be seen for larger energies. Different from aluminum, there are several roots of the real part of $\varepsilon_{\mathrm{Ag}}\left(\vec{q},\omega\right)$. The roots around $7\,\mathrm{eV}$, $22\,\mathrm{eV}$ and $29\,\mathrm{eV}$ result only from interband transitions and are nearly $\vec{q}$-independent for small $\vec{q}$-vectors. The two roots around $4\,\mathrm{eV}$ are, in contrast, influenced by intraband transitions und therefore $\vec{q}$-dependent. For small $\vec{q}$ these results are in good agreement with recent optical measurements and calculations,~\cite{Zhukov,Stahrenberg,Draxl2,Antonov} but in contrast to earlier measurements of Ehrenreich and Philipp,~\cite{Ehrenreich} where only the first root was found. Recent $GW$-calculations have yielded corrections to the DFT results for the band line-up~\cite{Onida} and have improved the agreement with experiment~\cite{Stahrenberg} further. For the purposes of determining \emph{the wave-vector dependent} characteristics, we neglect these small corrections. Figure \ref{fig:eps}(b) also shows the \emph{intraband} $\Re \varepsilon(\vec{q},\omega)$ of silver whose zero determines the intraband plasmon resonance.

Our numerical method allows us to get accurate results for the dielectric function even for low frequencies and small (but finite) wave vectors, where the behavior around $\omega =0$ is exclusively due to intraband transitions.~\cite{HaugSchmitt-Rink}  
Figure \ref{fig:eps2} shows the calculated low-frequency behavior of the imaginary part of the \emph{intraband} dielectric function for the simple metals aluminum and silver for different $q$. In contrast to the Drude peak at $q=0$, we find a double-peak structure even for small $q$ for both materials. This spectral signature results from the anisotropy of the bands, i.e.~the conduction bands cross the Fermi energy with different slopes. For smaller $q$ values the two peaks move closer together. Since for $q\to0$ they must merge into the Drude peak, we also show the result for $\varepsilon_2(q=0,\omega)$, as obtained from the ELK code, according to Refs.~\onlinecite{ELK,Draxl}. The $q=0$ results are calculated with a broadening $\hbar\gamma$ in~\eqref{eps} and we can draw some conclusions for $\gamma$ from our finite $q$ results. A broadening of $\hbar\gamma = 0.01$\,Ha at $q=0$ seems too large to compare well with our finite $q$ spectra, since the low-frequency resonance becomes narrower and higher with decreasing $q$. For the low frequency behavior of the dielectric function, our results suggest that one should use a broadening of at most $\hbar\gamma=0.005$\,Ha in aluminum and even less in silver. In contrast, for larger energies, where interband transitions play a role, $\hbar\gamma=0.005$\,Ha seems to be small enough for silver according to Ref.~\onlinecite{Zhukov}. This difference shows that the broadening should actually be energy-dependent in the calculation of the dielectric function.    


\begin{figure}
\includegraphics[width=0.40\textwidth]{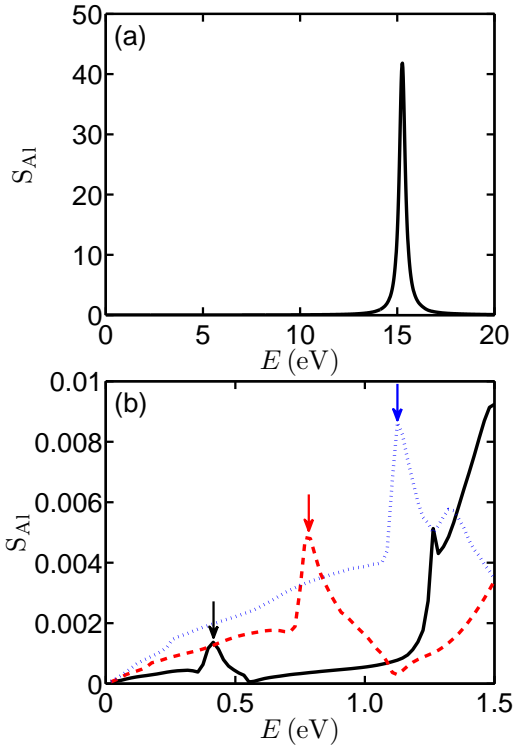}
\caption{\label{fig:lossAl}(Color online) Loss function of aluminum on a large (a) and a small energy scale (b). The arrows mark the peaks due to the acoustic plasmon resonance. The wave vectors are the same as in Fig.~\ref{fig:eps2}.} 
\end{figure}

\begin{figure}
\includegraphics[width=0.40\textwidth]{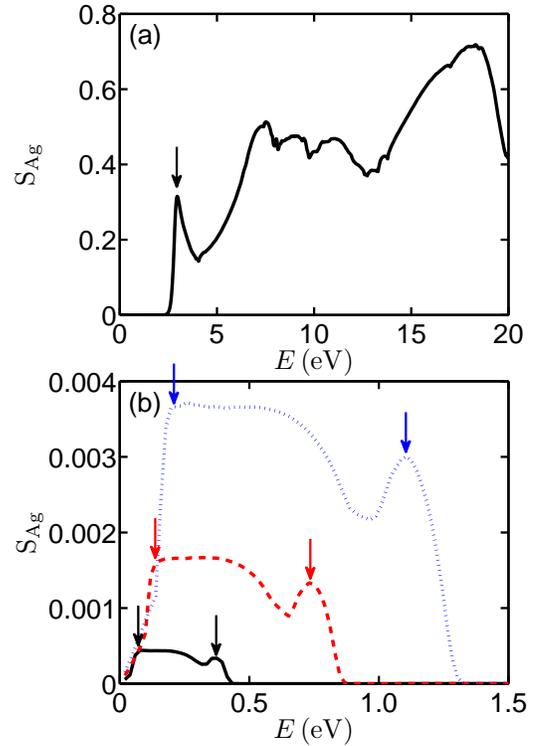}
\caption{\label{fig:lossAg}(Color online) Loss function of silver on a large (a) and a small energy scale (b). The arrow in figure (a) marks the optical plasmon and the arrows in figure (b) correspond to the edges of the acoustic plasmon signature. The wave vectors are the same as in Fig.~\ref{fig:eps2}.} 
\end{figure}

Having examined the numerical results for the dielectric function for long-wavelengths, we now turn to the properties of the electron plasma at finite $\vec{q}$. The conventional way to characterize plasmons is as a resonance of the loss function, or dynamic structure factor, $S=-\Im\varepsilon^{-1}(\vec{q},\omega)$,~\cite{Mahan,Hummel} which is shown for completeness in Figs.~\ref{fig:lossAl} and \ref{fig:lossAg}. In the electron-gas model it is known that $S$ exhibits a sharp peak at the plasma frequency~$\omega_{\mathrm{Pl}}(\vec{q})$, which dominates $S$ at small $q$ and whose dispersion is~$\omega_{\mathrm{Pl}}(\vec{q})\propto q^2$. This is the signature of a collective plasma excitation (or ``plasmon''). The loss function of aluminum (full calculation), Fig.~\ref{fig:lossAl}(a), shows a single plasmon peak at $\omega_{\mathrm{Pl}}=15\,\mathrm{eV}$. The loss function of silver, Fig.~\ref{fig:lossAg}(a), is also in agreement with earlier calculations,~\cite{Onida,Hummel,Echenique} where the first peak around $4\,\mathrm{eV}$ was identified as the ``optical plasmon''.

Recent investigations of the loss function of different materials yielded additional peaks at small energies, which were called ``acoustic plasmons''.~\cite{Silkin1,Silkin2,Silkin3,Silkin4} These peaks are due to the double-peak structure of the imaginary part of the dielectric function at small energies.~\cite{Silkin1,Silkin2,Silkin3,Silkin4} We show our results for the loss function of aluminum and silver for small energies in Figs.~\ref{fig:lossAl}(b) and \ref{fig:lossAg}(b), which demonstrate the existence of such acoustic plasmon excitations also in these materials. Aluminum has a well-defined acoustic plasmon resonance, whereas for silver we find a broad spectral signature instead of a sharp resonance. These different spectral shapes in the acoustic plasmon region are due to the different separations between the peaks in $\varepsilon_2$, cf.~Fig.~\ref{fig:eps2}(a) and (b).

\subsection{Plasma Frequencies and Plasmon Dispersions}

In this subsection we focus on intraband and full plasmon dispersions. The numerical value of both plasma frequencies can be expressed in the form $\hbar\omega_\mathrm{Pl}= \sqrt{ne^2/(\varepsilon_0m)}$.~\cite{HaugKoch} For the plasma frequency obtained from the full dielectric function one finds~\cite{Segall} 
\begin{equation}
\hbar\omega_{\mathrm{Pl}}(\mathrm{Al})=\sqrt{\frac{n^*e^2}{\varepsilon_0m_0}}\approx 15\,\mathrm{eV},
\end{equation}
where $m_0$ is the vacuum electron mass and $n^*$ is an effective electron density that is not equal to the valence electron density. For Al, typical results are $n^*=1.8\,\mathrm{e}/\mathrm{atom}\textrm{-}2.6\,\mathrm{e}/\mathrm{atom}$ (see Ref.~\onlinecite{Segall} and references therein). From the intraband dielectric function, one finds~\cite{Segall} 
\begin{equation}
\hbar\omega_{\mathrm{intra}}(\mathrm{Al})=\sqrt{\frac{n_\mathrm{e}e^2}{\varepsilon_0m_\mathrm{opt}}}\approx 12\,\mathrm{eV},
\label{intraband-plasma-density}
\end{equation}
where $m_\mathrm{opt}$ is an effective (``optical'') mass with typical values of $m_\mathrm{opt}=1.15\,m_0\textrm{-}1.67\,m_0$ (see Ref.~\onlinecite{Segall} and references therein) and $n_\mathrm{e}$ the density of conduction electrons in aluminum. More generally, the intraband plasma frequency tensor can be computed in ab-initio fashion,~\cite{Draxl}
\begin{equation}
\hbar^{2}\omega_{\mathrm{intra},ij}^{2}=\frac{\hbar^{2}e^{2}}{\pi m_{\mathrm{opt}}^{2}}\sum_{\mu}\int d^3k \,\langle p^{i} \rangle _{\mu\vec{k}}\langle  p^{j} \rangle _{\mu\vec{k}}\delta(\varepsilon_{\vec{k}}^{\mu}-E_{\mathrm{F}}),
\label{intraband-plasma}
\end{equation}
where $E_{\mathrm{F}}$ denotes the Fermi energy and $\langle p^{i} \rangle _{\mu\vec{k}}=\langle \psi^\mu_{\vec{k}}|p^{i}|\psi^{\mu}_{\vec{k}}\rangle$ the momentum matrix element between two Bloch states. It seems to be accepted wisdom that, for the purposes of comparison with electron loss spectroscopies, one should use the plasma frequency as given by the peak of the loss function for the \emph{full} dielectric function, even though such a well-defined peak in the loss function does not always exist. On the other hand, for the purpose of the description of the Drude peak in optical spectra one should use the intraband plasma frequency.~\cite{Draxl,Segall} The intraband plasma frequency also has the advantage, that it can be calculated ab-initio by Eq.~\eqref{intraband-plasma}, even if there is no well defined peak in the loss function obtained from the full dielectric function. Further, this plasma frequency is related to the model of an electron gas with the actual density of valence electrons and an effective optical mass. Thus, the relation between the plasma frequency and the (correct) electron density as mentioned above should only be used with the pure intraband calculation. 

If one takes the point of view that the plasma frequency is intimately connected with excitations in a \emph{single-band} electron gas, the intraband plasma frequency seems the most faithful generalization of the plasma frequency to a real material, whereas interband transitions are mixed in the full dielectric function, even in the limit of $q\to0$. Indeed, early investigations of Ehrenreich and Philipp~\cite{Ehrenreich} for silver used this point of view to reconcile an apparent discrepancy between the full plasma frequency and the intraband plasma frequency by subtracting the interband contributions from the full dielectric function. This analytic subtraction procedure for the dielectric function, and the separate calculation of the intraband plasma frequency according to Eq.~\eqref{intraband-plasma} or \eqref{intraband-plasma-density} works only at $q=0$. For finite $q$ one has to obtain the dielectric function first, and then extract the plasma frequency. 

To study the difference between these two points of view we calculate the plasma dispersion for finite wave vectors for both cases. For Al and Ag, the full dielectric function leads to a loss function with a well-defined peak that can be used to obtain the full plasmon energy for a range of $q$ values. In addition to these full plasmon dispersions, we also calculate the $q$-dependent \emph{intraband} plasmon energy, which we obtain from the root of the intraband $\Re \varepsilon(q,\omega)$. We use the zero of $\Re \varepsilon(\vec{q},\omega)$ (for fixed $\vec{q}$) instead of the peak in the intraband loss function because the intraband plasmon resonance does not have a finite broadening in the RPA. As support for the validity of this procedure we note that, for $q\to0$, it coincides with the intraband plasma frequency as calculated via Eq.~\eqref{intraband-plasma}.

The classification of intraband vs.~interband transition, which underlies our numerical results, is based on the labeling of the bands. We use a generic and unique band labeling according to the sequence of Kohn-Sham energy eigenvalues at each $\vec{k}$-point.~\cite{Lax}

\begin{figure}
\includegraphics[width=0.40\textwidth]{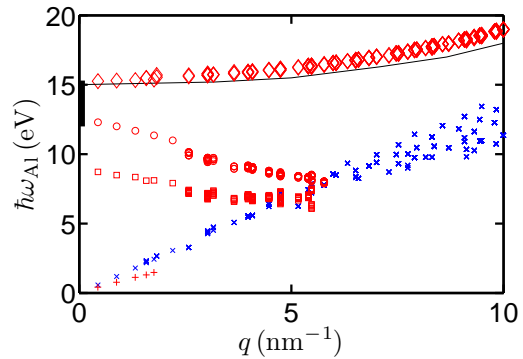}
\caption{\label{fig:eeconti_and_plasma_al}(Color online) Different plasmon dispersions (red markers) and top of the \emph{intraband} electron-hole continuum (blue crosses) of aluminum. Shown are the effective plasma frequency (diamonds), intraband plasma frequency (open circles) and intraband plasma frequency without transitions between Kramers pairs of states (open squares) using $17^3$ and $61^3$ $\vec{k}$-points in the 1.~BZ. The scatter of the continuum and the plasma frequencies is due to the different $\vec{q}$-vectors with the same modulus. The black bar at $q=0$ denotes the range of the published values for $\omega_{\mathrm{Pl}}\left(\vec{q}\rightarrow0\right)$~\cite{Zeman,Shiles,Kittel} and the thin solid line corresponds to the experimental results of Ref.~\onlinecite{Fink}. Acoustic plasmons in the $\mathrm{\Gamma L}$-direction are indicated by red ``+''.}
\end{figure}

In Fig.~\ref{fig:eeconti_and_plasma_al} we show a comprehensive plot of the dispersions of the different plasmons along with edge of the intraband continuum for Al. In particular, the (red) open circles are the \emph{intraband} plasma frequencies and the blue crosses mark the upper border of the \emph{intraband} electron-hole continuum and therefore describe the maximal energy transfer, $\underset{\vec{k}}{\mathrm{max}}\{\hbar\omega = \epsilon(\vec{k}+\vec{q})-\epsilon(\vec{k})\}$, for intraband single-particle transitions with a change in the wave-vector. Several markers at the same $q$-value correspond to different wave vectors $\vec{q}$ with the same modulus. The spread of values at a given $|\mathbf{k}|$-point therefore is a measure of the anisotropy of the spectrum. We also show as (red) open squares an ``extreme intraband'' calculation where we neglect transitions between Kramers degenerate bands in the intraband dielectric function. Finally, the plasma dispersion $\omega_\mathrm{Pl}$, determined from the peak of the full loss function is plotted as (red) diamonds. For small $q$, the full plasma dispersion reaches a value of $15\,\mathrm{eV}$, which is in agreement with results extracted from measurements, such as EELS, where all transitions contribute.~\cite{Fink,Shiles,Kittel} The intraband plasma frequency goes to $12\,\mathrm{eV}$, for $q\to0$, which is in agreement with values obtained from fits to electron-gas models, see, e.g., Ref.~\onlinecite{Zeman}, and corresponds to the result of Eq.~\eqref{intraband-plasma} obtained with the ELK-code.~\cite{ELK} 

The most striking result is a qualitative difference of the intraband plasmon dispersion from the electron-gas plasmon dispersion, where the plasma frequency \emph{increases} as $\omega_{\mathrm{Pl}}^{\text{e-gas}}\propto q^2$. The computed intraband plasma frequency \emph{decreases} until it joins the electron-hole continuum. It is obvious that the intraband plasma frequency has to curve downward, because in some $\vec{q}$-directions the high-energy boundary of the intraband electron-hole continuum is smaller than the (intraband) plasma frequency in the long wavelength limit, $q\rightarrow0$. This will become even more obvious in the case of silver below. Note that we get a similar behavior of the intraband plasma frequency by evaluating the f-sum rule (not shown here),~\cite{Vignale,Mahan}
\begin{equation}
\omega_{\mathrm{intra,f\textrm{-}sum}}^{2}(\vec{q})=\frac{2}{\pi}\int_{0}^{\infty}d\omega\,\omega\varepsilon_{2}(\vec{q},\omega),
\end{equation}
which is $\vec{q}$-independent for a single parabolic band.~\cite{Vignale} The negative dispersion thus clearly comes from a finite width of the bands, for which we take the intraband transitions into account. The decrease of the intraband Coulomb matrix elements for larger momenta $\mathbf{q}$, which originates from the Coulomb potential, leads to smaller contributions to $\varepsilon_2$ for large wave vectors and can therefore also have an influence on this negative dispersion, which was also mentioned in Ref.~\onlinecite{Silkin}.

Another remarkable property of the intraband plasma frequency is the pronounced contribution of transitions between Kramers degenerate bands in \eqref{eps}, which can be seen by comparing the plasma frequencies obtained from the intraband dielectric function calculated with and without these contributions. This is a consequence of the spin mixing in the single-particle states,~\cite{Fabian} because the matrix elements $B_{\vec{k}\vec{q}}^{\mu\nu}$ between Kramers degenerate bands would vanish if these were completely spin-polarized. Without spin-mixing, the inclusion of transitions between Kramers degenerate bands would make no difference. 

The full plasmon dispersion in Fig.~\ref{fig:eeconti_and_plasma_al} is in agreement with EELS measurements~\cite{Fink} and calculations.~\cite{Eguiluz,Lee} Note that the interband contributions reduce the spread for different $\vec{q}$ directions, i.e., the anisotropy, and, more importantly, change the overall shape of the dispersion qualitatively. That there is an influence of interband transitions on the effective plasma frequency was already noted by Quong and Eguiluz,~\cite{Eguiluz} but, these authors extracted effective dispersion parameters that described relatively small deviations from the plasmon of a gas of electrons with vacuum electron mass $m_0$. Compared with the intraband plasmon, however, the interband contributions lead to \emph{qualitative} changes.

\begin{figure}
\includegraphics[width=0.40\textwidth]{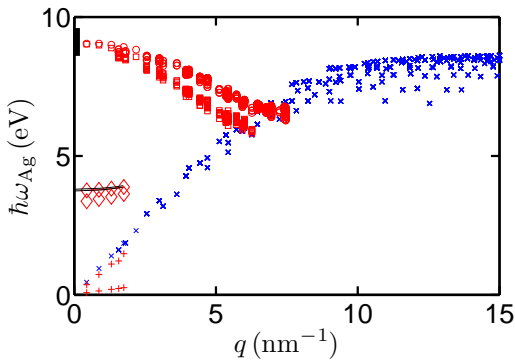}
\caption{\label{fig:eeconti_and_plasma_ag}(Color online) Same as Fig.~\ref{fig:eeconti_and_plasma_al} for silver. The first two roots of $\varepsilon_{1}(\vec{q},\omega)$ (full calculation) in the $\mathrm{\Gamma L}$-direction are plotted as open diamonds. These roots give rise to the optical plasmon resonance marked in Fig.~\ref{fig:lossAg}(a). Published theoretical results for $\omega_{\mathrm{Pl}}(\vec{q})$ are indicated for $q=0$ (black bar on the energy axis).~\cite{Zeman,Hooper,Noguez,Kreibig} The measured curves for $q>0$ from Refs.~\onlinecite{Piazza} and ~\onlinecite{Zacharias} (black lines) are indistinguishable on this energy scale.}
\end{figure}


Figure \ref{fig:eeconti_and_plasma_ag} displays the different plasmon dispersions and continuum edges for silver. We assign the effective plasma frequency to the closely spaced roots of $\varepsilon_{1}(\vec{q},\omega)$ at about $4\,\mathrm{eV}$ as shown in Fig.~\ref{fig:eps}(b). These are the ones that show a dispersion with $q$ and provide a dominant contribution to the loss function. In fact, the $GW$-RPA calculation~\cite{Onida} shows that the two roots at this energetic position are even closer and that the peak in the loss function is sharper than in our DFT-RPA calculation, so that the identification of this resonance with an ``optical plasmon'' is correct. The parabolic dispersion of the effective plasma frequency around $\omega_{\mathrm{Pl}}(\mathrm{Ag}) \approx 4\,\mathrm{eV}$ determined from the full dielectric function fits well to measurements by electroreflectance spectroscopy~\cite{Ehrenreich,Piazza} and EELS~\cite{Zacharias} and calculations.~\cite{Draxl2} For $q\to 0$ our intraband calculation, $\omega_{\mathrm{intra}}(\mathrm{Ag}) \approx 9\,\mathrm{eV}$, is again in good agreement with electron-gas models~\cite{Zeman,Hooper,Noguez,Kreibig} and the ELK-code.~\cite{ELK} That the full plasmon dispersion in Ag is markedly different from the one of Al due to $d$-electron contributions was already established by experiments~\cite{Piazza} and calculations.~\cite{Draxl2} Our results complete this picture by showing that the negative intraband plasmon dispersion is actually quite similar to the one of Al, except that the influence of spin mixing is not as pronounced. 

Thus, our calculations complete the picture of the different plasmon dispersions in these metals and make the electron-gas model questionable at finite $q$, even for Al, because both choices for the plasmon have some unappealing consequences. Using the plasma frequency $\omega_{\mathrm{Pl}}$ leads to a curvature of the plasmon dispersion as expected from the electron-gas model, but only because of interband transitions, which are outside of a single-band electron-gas model. The intraband plasmon, which is, in the optical limit, defined via Eq.~\eqref{intraband-plasma} and can be used to determine electron densities within the electron-gas model (Eq.~\eqref{intraband-plasma-density}),~\cite{Segall} shows a downward slope, which does not agree with the electron-gas plasmon dispersion. In our view, and as will be shown below, such a downward slope of the intraband plasmon dispersion is quite generic. It is due to the finite width of the bands, and not influenced much by \emph{details} of the band structure. The finite width of the bands, of course, is related to hybridization and band splitting in a real material, which is absent in the idealized electron-gas model with one parabolic band of effectively infinite width. 

For completeness we also show the dispersion of the acoustic plasmon resonance of aluminum in Fig.~\ref{fig:eeconti_and_plasma_al} and the edges of the acoustic plasmon signature of silver in Fig.~\ref{fig:eeconti_and_plasma_ag}. We find a linear dispersion of these resonances, which is the reason that they were called acoustic plasmons in Refs.~\onlinecite{Silkin1,Silkin2,Silkin3,Silkin4}.

With an understanding of these simple cases, we turn to an investigation of the plasmon behavior in more ``complicated'' materials.

\section{Heusler Compounds\label{resultsheusler}}

\subsection{Dielectric Function and Loss Function}

We investigate here the Heusler compounds $\mathrm{Co}_2\mathrm{Mn}\mathrm{Si}$ (CMS) and $\mathrm{Co}_2\mathrm{Fe}\mathrm{Si}$ (CFS). The calculations are done as described in the previous sections, with the difference that we employ a local-density appoximation (LDA+$U$) calculation with a Hubbard-$U$ in the fully localized limit within the ELK FP-LAPW code.~\cite{ELK} The crystal structure is a face-centered cubic lattice with lattice constants $a_{\mathrm{CMS}}=a_{\mathrm{CFS}}=5.654\,\textrm{\r{A}}$.~\cite{Webster,Balke} The basis of the lattice is formed by a $\mathrm{Mn}$/$\mathrm{Fe}$ atom located at $(0,0,0)$, cobalt atoms at $(\frac{1}{4},\frac{1}{4},\frac{1}{4})$ and $(\frac{3}{4},\frac{3}{4},\frac{3}{4})$, and a $\mathrm{Si}$-atom at $(\frac{1}{2},\frac{1}{2},\frac{1}{2})$, as outlined in Ref.~\onlinecite{Graf}. The effective screening parameter (Hubbard-$U$) is introduced to include interactions between electrons in narrow bands.~\cite{Minar} We choose $U_{\mathrm{Mn}}=0.195\,\mathrm{Ha}$, $U_{\mathrm{Co}}=0.15\,\mathrm{Ha}$ for CMS~\cite{Balke} and $U_{\mathrm{Co}}=0.155\,\mathrm{Ha}$ and $U_{\mathrm{Fe}}=0.16\,\mathrm{Ha}$ for CFS~\cite{Balke} to reproduce the density of states (DOS) calculated in Ref.~\onlinecite{Balke} with the ELK FP-LAPW code (not shown). Again, the $\vec{k}$- and band-resolved energies and distribution functions as well as the matrix elements between different states serve as input parameters for our evaluation of the dielectric function for all wave-vectors $\vec{q}$ and frequencies $\omega$. 

\begin{figure}
\includegraphics[width=0.40\textwidth]{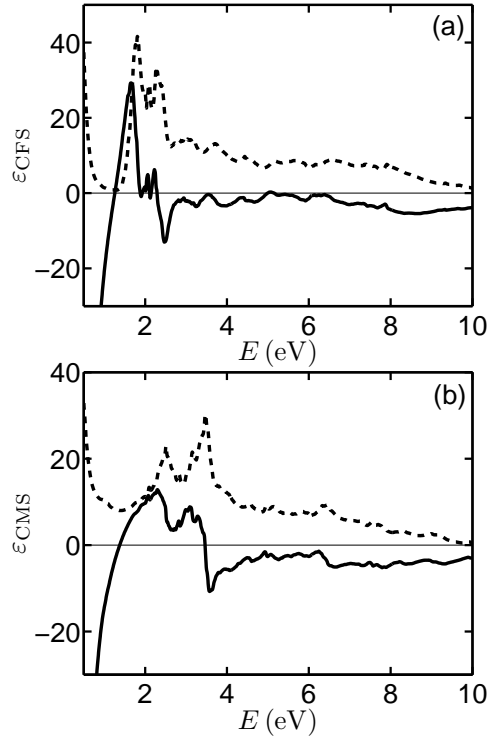}
\caption{\label{fig:eps_Heusler} Computed real (black solid line) and imaginary (black dashed line) part of the dielectric function of CFS (a) and CMS (b) for $q = 0.55\,\mathrm{nm}^{-1}$, where we used $35\times35\times35$ $\vec{k}$-points in the (full) 1.~BZ and included the first 62 (60) bands in the CFS (CMS) calculation.} 
\end{figure}

Figure \ref{fig:eps_Heusler}(a) shows our numerical results for the real and imaginary part of the dielectric function of CFS for $q = 0.55\,\mathrm{nm}^{-1}$. For small energies and momentum transfers, the strong increase of the imaginary part is again dominated by intraband transitions, which produce the Drude peak in the optical limit $(q \to 0)$. The effect of interband scattering processes can be seen for energies larger than $1.5\,\mathrm{eV}$. Although this curve is similar to the calculations of Picozzi et al.,~\cite{Picozzi} there are differences (e.g.~the positions of the roots of $\varepsilon_1$), which result mainly from our use of a finite $\vec{q}$ value and the inclusion of a larger number of bands, which also explains the small difference in the DOS.

In the dielectric function of CMS, shown in Fig.~\ref{fig:eps_Heusler}(b), the intraband and interband contributions are not clearly separated, but the overall shape is very similar to $\varepsilon_{\mathrm{CFS}}$ over a broad energy range. The result is in good agreement with recent $GW$ calculations that yield only slightly different energies.~\cite{Meinert}  

\begin{figure}
\includegraphics[width=0.40\textwidth]{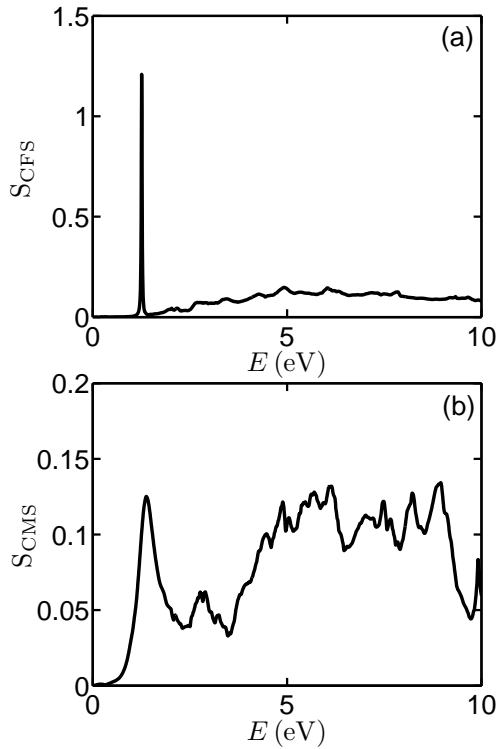}
\caption{\label{fig:loss_Heusler} Computed loss function for CFS (a) and CMS (b) for the parameters given in Fig.~\ref{fig:eps_Heusler}.} 
\end{figure}

\begin{figure}
\includegraphics[width=0.40\textwidth]{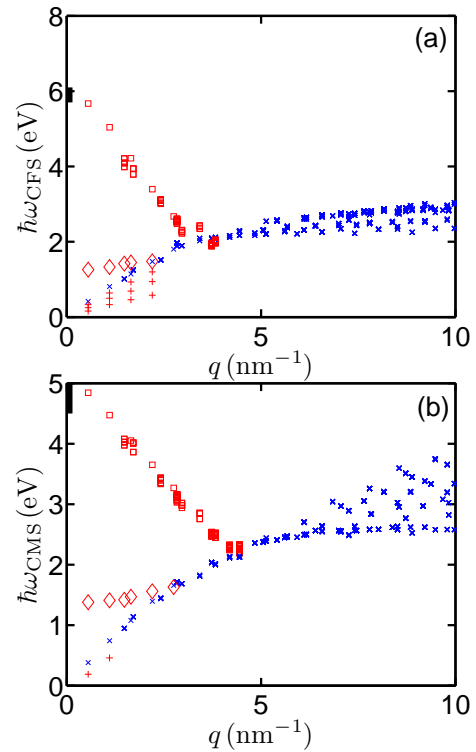}
\caption{\label{fig:plasma_Heusler}(Color online) Same as Fig.~\ref{fig:eeconti_and_plasma_ag} for CFS (a) and CMS (b) using $13^3$ and $35^3$ $\vec{k}$-points in the (full) 1.~BZ. The black bars at $q=0$ correspond to the range of values obtained by the DFT-code\cite{ELK} and recent calculations.\cite{Picozzi,Meinert,Kumar}} 
\end{figure}

The first root of the real part of $\varepsilon_{\mathrm{CFS}}$ coincides with a nearly vanishing imaginary part at about $1.3\,\mathrm{eV}$, which results in a strong peak in the loss function and therefore a well defined plasmon energy $\omega_{\mathrm{Pl}}$, as shown in Fig.~\ref{fig:loss_Heusler}(a) and already discussed in the case of silver. In contrast, intraband plasma frequency calculations determined via Eq.~\eqref{intraband-plasma} by standard DFT-codes~\cite{ELK} and recent calculations~\cite{Kumar,Meinert} for CFS lie between $5.7\,\mathrm{eV}$ and $6.1\,\mathrm{eV}$. In Fig.~\ref{fig:loss_Heusler}(b), the first peak in the loss function of CMS at about $1.4\,\mathrm{eV}$ is also an optical plasmon. Even though the resonance is not as pronounced as in the case of CFS, it can still be well distinguished from the quasi-continuous spectrum. For CMS, the intraband plasma frequency ~\eqref{intraband-plasma}, as determined by standard DFT-codes~\cite{ELK} and in previous calculations~\cite{Picozzi,Meinert} is in the range of $4.5\,\mathrm{eV}\textrm{-}5.0\,\mathrm{eV}$, which has no counterpart in the loss function shown in Fig.~\ref{fig:loss_Heusler}(b).  

As in the case of the simple metals aluminum and silver, we determine the dispersions of the full plasmons, i.e., the one related to peaks in the loss function, and the ``intraband plasmon''. Due to their high spin polarization around the Fermi energy, the two Heusler compounds show no Kramers degeneracy, and we therefore need not distinguish between different intraband scattering processes. 

\subsection{Plasmon Dispersions}

In Fig.~\ref{fig:plasma_Heusler} we show our results for the full and intraband plasmon dispersions of CFS and CMS. For both materials, the optical limit of the pure intraband calculation (red squares) reproduces the intraband plasma frequency calculated from Eq.~\eqref{intraband-plasma}.~\cite{ELK} As in the case of the simple metals the intraband plasmon shows a negative dispersion due to the finite width of the conduction bands. The full plasmon, i.e., peak of the loss function, is a few eVs away and moves to higher energies for larger momenta $q$. The behavior of the different plasma frequencies is in many aspects similar to the case of silver, regardless of the much more complicated band structure of the Heusler compounds. Note also that both the full and the intraband plasmon dispersions get heavily damped when they reach the intraband electron-hole continuum. Another difference between the two Heusler compounds can be seen in terms of the acoustic plasmons indicated as ``+'' in Fig.~\ref{fig:plasma_Heusler}. In CFS we find three dispersive acoustic plasmons in the small energy regime, but in CMS only one, which descends in the total loss function spectrum even for small wave vectors.

Although the differences between the plasma frequencies have been known at $q=0$, we believe that the momentum dependent dispersions for both simple metals and complicated Heusler compounds, show the distinction between the two different plasmon dispersions much more clearly: one increases and one decreases with $q$. We stress that there does not seem to be a clear candidate for ``the'' plasmon. Thus, when using plasmon energies as input for calculations or fitting measurements by plasmon energies, one should clearly distinguish between intraband and the full plasmon.

\section{Conclusions\label{conclusions}}

In conclusion, we presented a comprehensive study of plasmon dispersions in simple metals and Heusler compounds based on an accurate ab-initio evaluation of the RPA dielectric function. The dynamical dielectric functions were evaluated for all vector momenta $\vec{q}$ and frequencies $\omega$ from DFT-input using a wave vector-dependent linear tetrahedron method. The dispersion of the plasma frequency was obtained from the full dielectric function and from a calculation including only intraband transitions. We found that the plasma frequency derived from the intraband dielectric function agrees for $q\to0$ with the intraband plasma frequency obtainable from standard DFT codes, but shows a remarkable negative dispersion. Based on these results, it was argued that neither the full nor the intraband plasmon can be put into correspondence with the single-band electron-gas model at finite $q$ without problems: The full dielectric function always contains the influence of interband transitions, and does not always yield a clear resonance, whereas the intraband plasma frequency yields a plasmon dispersion with a downward curvature. From the intraband plasmon dispersion it was also shown that there is a pronounced spin mixing in aluminum. For the bulk Heusler compounds CFS and CMS we presented spectra of dielectric and loss functions in the optical limit and obtained well defined plasmon resonances from the full dielectric function. We found that these are actually ``optical'' plasmon resonances dominated by interband transitions. For the intraband plasmon, on the other hand, we found a qualitative behavior that is quite similar for CFS and CMS even though their band structures are very different. The intraband plasmon properties are not much different from those of the simple metals aluminum and silver: They also show a downward curvature. Furthermore, we found signatures of acoustic plasmons in all materials. We stressed that our results provide, in general, a clear qualitative differentiation between the ``full'' and intraband plasmons at finite $q$ vectors that may be blurred at $q=0$, where, for instance, in aluminum, the two plasmons may appear very similar.

\section*{Acknowledgements}
We are grateful to the J\"{u}lich Supercomputer Centre (JSC) for a CPU-time grant. We have benefitted from discussions with C.~Draxl (Berlin), W.~H\"{u}bner (Kaiserslautern), and P.~H\"{a}ussler (Chemnitz).

\end{document}